# Valley splitting in the transition-metal dichalcogenides monolayer via atoms adsorption

Xiaofang Chen[1], Liangshuai Zhong[1], Xiao Li[2*], Jingshan Qi[1*]

[1]*School of Physics and Electronic Engineering, Jiangsu Normal University, Xuzhou 221116, People's Republic of China*

[2]*Department of Physics, University of Texas at Austin, Austin, TX 78712, USA*

## Abstract

In this letter we study the valley degeneracy splitting of the transition-metal dichalcogenides monolayer by first-principles calculations. The local magnetic moments are introduced into the system when the transition-metal atoms are adsorbed to the monolayer surface. Zeeman effect arising from the local magnetic moment at transition-metal atom sites lifts the valley degeneracy. Anomalous charge, spin and valley Hall Effects can be accessed due to valley splitting when we can only excite carriers of one valley. The valley splitting depends on the direction of magnetization and thus can be tuned continuously by an external magnetic field. This tunable valley splitting offers a practical avenue for exploring device paradigms based on the spin and valley degrees of freedom.

# I. INTRODUCTION

To date, electronic and spintronic devices[1] invariably rely on the fact that carriers carry electric charge and spin degrees of freedom, respectively. However, other degrees of freedom of carriers are urgently needed to broaden the view of novel devices design. The electronic structure of certain crystal lattices can contain multiple degenerate valleys, namely, degenerate conduction or valence band extrema in momentum space. The valley degree of freedom has enormous potential use in the development of valleytronic devices.[2-12] Transition-metal dichalcogenides $MX_2$ (M = Mo, W and X=S, Se and Te) monolayer harbors a pair of inequivalent degeneracy energy valleys in the vicinities of the vertices of a hexagonal Brillouin zone of the momentum space.[13] The principal challenge in the development of valleytronics is the ability to generate and manipulate the valley polarization in a controllable way. Dynamical polarization of monolayer $MX_2$ has been achieved by circularly polarized optical pumping,[14-16] offering a new paradigm for optoelectronic devices. However, dynamical valley polarization based on optical pumping relies upon creating a transient non-equilibrium photo-carrier distribution in which two valleys are populated differently, and is ultimately limited by carrier life times which can be quite short for $MX_2$.

An alternative route to valley polarization is to lift the valley degeneracy by breaking time-reversal symmetry. Valley splitting in $MX_2$ has been assessed in a few recent experiments[17-20], which showed that only small valley splitting, 0.1-0.2 meV/tesla, can be generated by an external magnetic field. Recent studies showed that the proximity-induced Zeeman effect is a more effective strategy to create larger valley splitting.[21-23] Zeeman field can be introduced by the magnetic substrate proximity effect or transition-metal (TM) atoms doping. However the magnetic substrate or TM atoms must be selected carefully to preserve the valley characters because the hybridization between the host $MX_2$ and the external substrate/doping, may destroy valley character. In this paper we take monolayer $MoS_2$ as an example to show that, the valley degeneracy can be lifted and valley characters can be well held when Sc, Mn, Fe and Cu atoms are adsorbed on the surface, providing an alternative way for controlling valley polarization.

## II. MODEL ANALYSIS

In its bulk form, $MX_2$ has the 2*H* stacking order with the space group $D_{6h}^4$, which is inversion symmetric. When it is thinned down to a monolayer, as shown in Fig. 1 (a), the crystal symmetry reduces to $D_{3h}^1$, and inversion symmetry is explicitly broken. In addition, $MX_2$ has a strong spin-orbit coupling(SOC) originated from the *d* orbitals of the heavy metal atoms.[24] For the monolayer $MX_2$, the inversion symmetry breaking together with strong SOC lifts the spin degeneracy of energy bands and makes it have peculiar coupled spin and valley physics[13], as shown in Fig. 1 (c). A large spin splitting (~150-500meV) appears at the highest valence bands(HVB) with opposite spin moments in two valleys, while a small spin splitting (~3-60meV) at the lowest conduction bands(LCB), owing to different orbital contributions between HVB and LCB, that is, the LCB state is mainly made of the M-$d_{z^2}$ orbital with small components from the M-$d_{xz}/d_{yz}$ and X-$p_x/p_y$ orbitals, while HVB state is mainly made of the M-$d_{x^2-y^2}/d_{xy}$ orbitals with some mixings from the X-$p_x/p_y$ orbitals. Besides, time-reversal symmetry requires that the spin splitting in opposite valleys must be opposite. Therefore, the HVB/LCB is still valley degenerate when SOC is considered. The opposite Berry curvature and spin moment in opposite valleys gives rise to both the valley and spin Hall Effect, but the vanished charge Hall Effect.[13]

When TM atoms are adsorbed to $MX_2$ monolayer surface, the local magnetic moments are introduced into $MX_2$. An effective Zeeman effect arises from the local magnetic moment at TM atom site, which lifts the degeneracy of two valleys with spin splitting. The Zeeman interaction can be expressed by a tight-binding term $H_{ex} = M \sum_{i,\alpha} c_{i\alpha}^+ \sigma_z c_{i\alpha}$, where $c_{i\alpha}^+$ creates an electron at TM atom site with spin $\alpha$, $\sigma_z$ is the $z$-component of spin Pauli matrices and $M$ describes the Zeeman field strength, which determines the valley splitting, $\Delta_{KK'} = E_K^{\downarrow} - E_{K'}^{\uparrow}$, an energy difference between two electronic states related by time-reversal operation. The opposite Zeeman field, denoted by the sign of $M$, leads to the opposite valley splitting. If $M$ is negative, $\Delta_{KK'}$ is positive; otherwise, if $M$ is positive, $\Delta_{KK'}$ is negative, as shown in Fig. 1 (d) and (e). Our studies show that for Mn and Fe adsorption cases $M$ is positive while $M$ is negative for Sc and Cu adsorption cases, indicating the different exchange coupling. In addition, according to this model we can also change $\Delta_{KK'}$ from positive to negative by reversing the direction of magnetization, as shown in Fig. 1 (e) and (f).

## III. COMPUTATIONAL METHOD

In the following, we take monolayer $MoS_2$ as an example to show our detailed results from first-principles calculation. We study the diluted TM atoms adsorption on the top of the monolayer $MoS_2$ by constructing a 4×4 supercell. Density functional theory calculations were performed in a supercell configuration using the VASP computer code.[25] We employed projector augmented-wave method for ion-electron interaction and the generalized gradient approximation of exchange-correlation functional.[26] We used an energy cutoff of 400 eV for the plane-wave basis. The supercell was adjusted to maintain a sufficiently large separation (20 Å) between adjacent monolayer. The two dimensional Brillouin zone integrations were carried out using 9×9×1 Monkhorst-Pack grids. We also tested more dense *k*-points set and higher cutoff energy and didn't find significant difference. Atomic coordinates within the supercell were fully optimized without any symmetry constraints with a criterion of the maximum residual force less than 0.01 eV/Å. For including strong correlation effects we performed GGA+U calculation[27] with a moderate effective $U_{eff}$ = (U-J)=4.0 eV for TM atoms.[28] We also check different $U_{eff}$ values, 2 and 6 eV. $U_{eff}$ can change the details of band structures, even magnetic moments and exchange coupling. Because it is difficult to evaluate $U_{eff}$ theoretically, we also discuss the effect of different $U_{eff}$ on valley character in this paper.

## IV. RESULTS AND DISCUSSION

### A. Adsorption Energy

Firstly by calculating the adsorption energy of three typical adsorption sites shown in Fig. 2 (a), we find that the most stable adsorption site is the top on Mo atom ($T_{Mo}$). There have been several works on the transition metal atom adsorption on monolayer $MoS_2$.[29-31] There are two main differences in our work: (1) We include strong correlation effects by GGA+U calculation for 3$d$ transition metal atoms because it is well known that the correlation effects is important for 3$d$ orbitals.[28] We find that the most stable adsorption site is the top on Mo atom for all 3$d$ transition metal atoms adsorption (Sc-Cu) by GGA+U method, while a hollow site is preferred for Sc, Mn and Fe adsorption by GGA calculation. However in these previous works only GGA method is used, leading to preferred hollow sites for Sc and Mn adsorption, which is in consistent with our GGA results. Therefore we think that including the strong correlation effects by GGA+U should be necessary for determining the preferred sites and more reasonable electronic structures. (2) Our work focuses on the concept of the valley polarization in $MoS_2$ by the transition metal atoms adsorption, which is not discussed by other papers.

In the following, we focus on this most stable adsorption configurations. The calculated adsorption energy and magnetic moments are listed in Table1. For Sc, Ti, V, Co, Ni and Cu, the adsorption energy is large, indicating a strong hybridization, while the adsorption energy is relatively small for Cr and Mn, indicating a moderately weak adsorption. For Sc, Ti and V, strong hybridization make the 4$s$ electrons transfer to the unoccupied 3$d$ orbital. Since the 3$d$ shell is less than half filled, all 3$d$ electrons possess the parallel spins according to Hund's rule, leading to the increase of the magnetic moment. For Cr, Mn and Fe, the hybridization doesn't largely affect the 3$d$-orbital energy and thus all 3$d$ electrons align spin polarization according to the Hund's rule. Due to the magnetic proximity effect Cr atom strongly magnetizes the surrounding electrons, which increases the total magnetic moment from 5.0 to 6.0 $\mu_B$. For Ni, the strong hybridization lowers the 3$d$-orbital energy and makes two electrons transfer from the 4$s$-orbital to the 3$d$-orbital, which forms a closed 3$d$ shell with a vanishing magnetic moment. For Cu, because of the closed $d$ shell, the strong hybridization makes electron partially transfer from the Cu 4$s$-orbital to the Mo 4$d_{z^2}$-orbital, which together contributes to the magnetic moment of 1.0 $\mu_B$.

## B. Valley Splitting

Then we study the valley splitting by calculating the band structures for pristine and TM atoms adsorbed MoS$_2$ monolayer. A 4×4 supercell of the monolayer MoS$_2$ is shown in Fig. 2 (a) and the corresponding reciprocal momentum space structures and high symmetry points are shown in Fig. 2 (b). Band structure of a 4×4 supercell of the pristine monolayer MoS$_2$ is shown in Fig. 2 (c). As mentioned before, it has two inequivalent valleys in the momentum space electronic structure for low energy carriers, which localize at the corners (K and K' points) of 2D hexagonal Brillouin zone. Inversion symmetry breaking together with strong SOC leads to the spin splitting and this spin splitting is opposite for two valleys due to the time-reversal symmetry. When TM atoms are adsorbed to MoS$_2$ monolayer surface, the local magnetic moments are induced into MoS$_2$. Band structures with SOC are shown in Fig. 3 when the spin quantization axis is perpendicular to monolayer plane. Exchange interaction between the local magnetic moment at TM site and the spin moment of valley electrons indeed lifts the valley degeneracy. For Sc, Mn, Fe and Cu adsorption, the spin splitting of HVBs becomes different for two inequivalent valleys, leading to the valley splitting (HVB for one valley is higher than that for the other valley). For Ti, V, Cr and Co adsorption cases, TM impurity bands hybrid with HVBs of the host $MoS_2$, leading to indistinguishable valleys. In Fig. 3, the red curves are used to denote the highest valence bands of the host $MoS_2$ and $\Delta_{KK'}$ denotes the valley splitting, as listed in Table 1.

As shown in Fig. 3, the spin splitting at K point is larger than that at K' point for Sc and Cu adsorption; thus the highest valence bands at K point is higher than that at K' point. Magnification of the valence bands around K and K' point is shown in the insertion of Fig. 3(a) for Sc adsorption. While, the spin splitting at K point is smaller than that at K' point for Mn and Fe adsorption and thus the highest valence bands at K point is lower than that at K' point. So, the valley splitting, $\Delta_{KK'}$, is negative for Mn and Fe adsorption, while is positive for Sc and Cu adsorptions. This origins from the opposite exchange interaction, positive *M* for Mn and Fe adsorption case and negative *M* for Sc and Cu adsorption. In addition we also show the field-tuning of the valley splitting by calculating $\Delta_{KK'}$ of Mn absorbed $MoS_2$ monolayer as the magnetization direction rotates from *x* direction(in plane) to *z* direction (perpendicular to monolayer plane). Fig. 4(a) demonstrates that $\Delta_{KK'}$ can be continuously tuned by rotating its magnetization direction. More importantly we can change $\Delta_{KK'}$ from positive to negative by reversing the direction of magnetization from +*z* to -*z* direction. So, the valley splitting can be easily controlled by an external magnetic field, a readily accessible experiment knob.

**C. Anomalous Charge/Spin/Valley Hall Effect**

A few interesting experiments immediately become compelling, to probe and manipulate the valley polarization. The berry curvature drives an anomalous transverse velocity in the presence of an electric field, $v_a \sim E \times \Omega(k)$, which is responsible for the intrinsic contribution to the anomalous Hall Effect.[32] Here, $\Omega(k)$ is the Berry curvature of Bloch electron, and $E$ is the applied electric field. For the pristine $MoS_2$ monolayer, the charge carriers in two valleys have opposite transverse velocities due to the valley degeneracy and opposite signs of the berry curvatures and thus the total anomalous Hall conductivity vanishes.[13] However, in TM adsorption systems the valley degeneracy is lifted and thus HVBs possess different energy for two opposite valleys. Due to the valley splitting, we can selectively dope only one valley with carriers, e.g. hole doping at K valley for Sc adsorbed $MoS_2$ monolayer as shown in Fig. 3 (a). Under an in-plane electric field the spin-up holes will produce a net transverse charge/spin/valley current, due to the Berry curvature driving anomalous velocity of Bloch electrons. This is an anomalous charge/spin/valley Hall effect, as shown in the insertion of Fig. 4(a). When we reverse the magnetization direction by an external magnetic field, an opposite directional charge/spin/valley Hall current is produced. Therefore we can control spin/valley polarization by carrier doping and an external magnetic field and detect them by the transversal voltage, which is crucial for spintronics and valleytronics.

For proving the anomalous charge, spin, and valley Hall effects, we further calculate $\Omega(k)$ since it is desirable to have sizable values especially near valleys of of MoS$_2$. For example, in Fig.4 (b), we show the calculated $\Omega(k)$ for a 4×4 supercell with an adsorbed Mn atom. It is obvious that $\Omega(k)$ is sharply peaked in the valley region and opposite signs for K and K'. Clearly, the valley identity remains intact apart from the valley splitting and shall display pronounced anomalous Hall effects.

At last, we show the results with different effective U$_{eff}$. For example, the band structures are shown in Fig.5 with U$_{eff}$ = 2 and 6 eV for Sc, Mn, Fe and Cu adsorption cases. The details of bands are changed for all cases. The most obvious change occurs for Sc adsorption case. When U$_{eff}$ increases to 6 eV, the sign of $\Delta_{KK'}$ change from positive to negative, indicating the different exchange coupling. However, for valley polarization application the most crucial issue is the magnitude of $\Delta_{KK'}$ because the sign of $\Delta_{KK'}$ can be tuned by an external magnetic field. For Sc, Mn and Cu adsorption case, the magnitude of $\Delta_{KK'}$ is not sensitive to different U$_{eff}$. For Fe adsorption case, $\Delta_{KK'}$ is -23, -28 and -66 meV when U$_{eff}$ is set to be 2, 4 and 6 eV, respectively. Although U$_{eff}$ affects the details of band structures and magnetic properties, the valley identity remains intact and the valley splittingis still preserved, which are crucial for valleytronic applications.

## V. CONCLUSION

In conclusion, we study the valley splitting in MoS$_2$ monolayer when TM atoms are adsorbed to the monolayer surface. For Sc, Mn, Fe and Cu adsorption, exchange interaction between the local magnetic moment at TM sites and the spin moments of valley electrons lifts the valley degeneracy. For Ti, V, Cr and Co adsorption case, TM impurity bands hybrid with the HVBs of the host $MoS_2$, leading to indistinguishable valleys. Besides, the valley splitting is also continuously tunable by using an external magnetic field to rotate the magnetization direction.Under an in-plane electric field a net transverse charge/spin/valley Hall current is produced by carrier doping due to the Berry curvature driving anomalous velocity of Bloch electrons and can be controlled by an external magnetic field. Therefore the tunable valley splitting adds a readily accessible dimension to the exploration of novel electronic degrees of freedom.

**AUTHOR INFORMATION**

**Corresponding Authors**

*E-mail: qijingshan@jsnu.edu.cn (J.Q), lixiao150@gmail.com (X.L)

**Notes**

The authors declare no competing financial interest.

**ACKNOWLEDGMENTS**

We acknowledge financial support from the National Natural Science Foundation of China (Projects No. 11204110, 11347005), PAPD and Open Fund of Key Laboratory for Intelligent Nano Materials and Devices of the Ministry of Education No. INMD-2015M04

**REFERENCES:**

[1] I. Žutić, J. Fabian and S. Das Sarma, Reviews of Modern Physics **76** (2), 323-410 (2004).

[2] A. Rycerz, J. Tworzydlo and C. W. J. Beenakker, Nature Physics **3** (3), 172-175 (2007).

[3] D. Xiao, W. Yao and Q. Niu, Physical Review Letters **99** (23), 236809 (2007).

[4] W. Yao, D. Xiao and Q. Niu, Physical Review B **77** (23), 235406 (2008).

[5] D. Gunlycke and C. T. White, Physical Review Letters **106** (13), 136806 (2011).

[6] Z. Wu, F. Zhai, F. M. Peeters, H. Q. Xu and K. Chang, Physical Review Letters **106** (17), 176802 (2011).

[7] J. Isberg, M. Gabrysch, J. Hammersberg, S. Majdi, K. K. Kovi and D. J. Twitchen, Nature Materials **12** (8), 760-764 (2013).

[8] Y. Jiang, T. Low, K. Chang, M. I. Katsnelson and F. Guinea, Physical Review Letters **110** (4), 046601 (2013).

[9] X. Li, T. Cao, Q. Niu, J. Shi and J. Feng, Proceedings of the National Academy of Sciences **110** (10), 3738-3742 (2013).

[10] Y. Liu, J. Song, Y. Li, Y. Liu and Q.-f. Sun, Physical Review B **87** (19), 195445 (2013).

[11] L. Ju, Z. Shi, N. Nair, Y. Lv, C. Jin, J. Velasco Jr, C. Ojeda-Aristizabal, H. A. Bechtel, M. C. Martin, A. Zettl, J. Analytis and F. Wang, Nature **520** (7549), 650-655 (2015).

[12] K. F. Mak, K. L. McGill, J. Park and P. L. McEuen, Science **344** (6191), 1489-1492 (2014).

[13] D. Xiao, G.-B. Liu, W. Feng, X. Xu and W. Yao, Physical Review Letters **108** (19), 196802 (2012).

[14] T. Cao, G. Wang, W. Han, H. Ye, C. Zhu, J. Shi, Q. Niu, P. Tan, E. Wang, B. Liu and J. Feng, Nature Communications **3**, 887 (2012).

[15] K. F. Mak, K. He, J. Shan and T. F. Heinz, Nat Nanotechnol **7** (8), 494-498 (2012).

[16] H. Zeng, J. Dai, W. Yao, D. Xiao and X. Cui, Nat Nanotechnology **7** (8), 490-493 (2012).

[17] Y. Li, J. Ludwig, T. Low, A. Chernikov, X. Cui, G. Arefe, Y. D. Kim, A. M. van der Zande, A. Rigosi, H. M. Hill, S. H. Kim, J. Hone, Z. Li, D. Smirnov and T. F. Heinz, Physical Review Letters **113** (26), 266804 (2014).

[18] G. Aivazian, Z. Gong, A. M. Jones, R.-L. Chu, J. Yan, D. G. Mandrus, C. Zhang, D. Cobden, W. Yao and X. Xu, Nature Physics **11** (2), 148-152 (2015).

[19] A. Srivastava, M. Sidler, A. V. Allain, D. S. Lembke, A. Kis and A. Imamoglu, Nature Physics **11** (2), 141-147 (2015).

[20] D. MacNeill, C. Heikes, K. F. Mak, Z. Anderson, A. Kormányos, V. Zólyomi, J. Park and D. C. Ralph, Physical Review Letters **114** (3), 037401 (2015).

[21] J. Qi, X. Li, Q. Niu and J. Feng, Physical Review B **92** (12), 121403 (2015).

[22] Q. Zhang, S. A. Yang, W. Mi, Y. Cheng and U. Schwingenschl?gl, Adv Mater **28** (5), 959-966 (2016).

[23] Y. C. Cheng, Q. Y. Zhang and U. Schwingenschlögl, Physical Review B **89** (15), 155429 (2014).

[24] Z. Y. Zhu, Y. C. Cheng and U. Schwingenschlögl, Physical Review B **84** (15), 153402 (2011).

[25] G. Kresse and J. Furthmuller, Computational Materials Science **6** (1), 15-50 (1996).

[26] J. P. Perdew, K. Burke and M. Ernzerhof, Physical Review Letters **77** (18), 3865-3868 (1996).

[27] S. L. Dudarev, G. A. Botton, S. Y. Savrasov, C. J. Humphreys and A. P. Sutton, Physical Review B **57** (3), 1505-1509 (1998).

[28] L. Wang, T. Maxisch and G. Ceder, Physical Review B **73** (19), 195107 (2006).

[29] J. Chang, S. Larentis, E. Tutuc, L. F. Register and S. K. Banerjee, Applied Physics Letters **104** (14), 141603 (2014).

[30]C. Ataca and S. Ciraci, Journal of Physical Chemistry C **115** (27), 13303-13311 (2011).

[31]Y. Z. Wang, B. L. Wang, R. Huang, B. L. Gao, F. J. Kong and Q. F. Zhang, Physica E **63**, 276-282 (2014).

[32]D. Xiao, M.-C. Chang and Q. Niu, Reviews of Modern Physics **82** (3), 1959-2007 (2010).

|  |  | Sc | Ti | V | Cr | Mn | Fe | Co | Ni | Cu |
|---|---|---|---|---|---|---|---|---|---|---|
| 4×4 supercell | Site | $T_{Mo}$ | $T_{Mo}$ | $T_{Mo}$ | $T_{Mo}$ | $T_{Mo}$ | $T_{Mo}$ | $T_{Mo}$ | $T_{Mo}$ | $T_{Mo}$ |
|  | $E_{ads}$ (eV) | 0.90 | 1.47 | 1.00 | 0.71 | 0.41 | 0.91 | 1.16 | 2.21 | 1.12 |
|  | $M_\mu$ ($\mu_B$) | 3.0 | 4.0 | 5.0 | 6.0 | 5.0 | 4.0 | 1.0 | 0.0 | 1.0 |
|  | $\Delta_{KK'}$ (meV) | 41 | - | - | - | -16 | -28 | - | - | 21 |

Table 1. Properties of nine 3$d$ transition metal atoms located at the lowest energy adsorption sites ($T_{Mo}$) on the 4×4 supercells of MoS$_2$ monolayer. The properties listed include the adsorption energy ($E_{ads}$), total magnetic moment of system ($M_\mu$), and the valley splitting ($\Delta_{KK'}$).

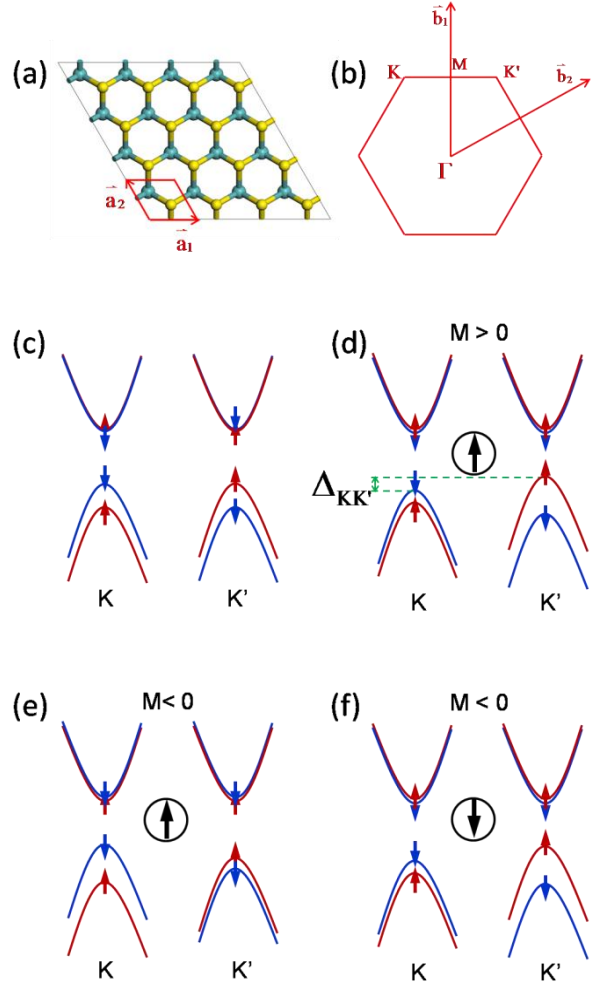

Figure.1. (a) Top view of the MX$_2$ monolayer. The red lines indicate the unit cell in plane. $\vec{a}_1$ and $\vec{a}_2$ denote the primitive vectors for the unit cell. (b) Corresponding reciprocal momentum space structures and high symmetry points. $\vec{b}_1$ and $\vec{b}_2$ denote the reciprocal primitive vectors for the unit cell. $\Gamma$, M, K and K$^{'}$ are the high symmetry points. (c), (d), (e) and (f) Schematic drawing of the band edge structures at two valleys. Down (up) opening parabolas are used to denote the highest valence (lowest conduction) bands. Red up-arrow and bluedown-arrow denote respectively spin-up and spin-down electron. The black arrows in (d), (e) and (f) denote the introduced local magnetic moment. $M$ is the exchange coupling constant in the interaction model Hamiltonian and $\Delta_{KK'}$ denotes the valley splitting.

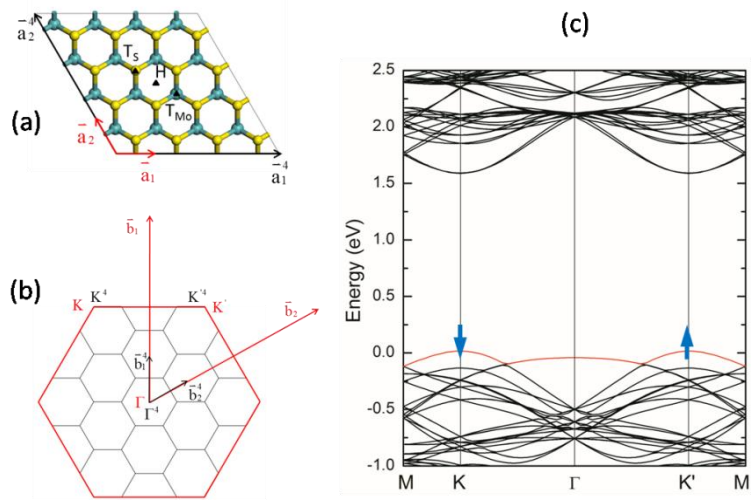

Figure.2. (a) A 4×4 supercell of the monolayer MoS$_2$. $\vec{a}_1^4$ ($\vec{a}_1$) and $\vec{a}_2^4$ ($\vec{a}_2$) denote the primitive vectors for the 4×4 (1×1) supercell. Three typical adsorption sites of single adatom on the monolayer MoS$_2$ are labeled as follows: hollow (H), top on Mo atom (T$_{Mo}$), and top on S atom (T$_S$). (b) Corresponding reciprocal momentum space structures and high symmetry points. $\vec{b}_1^4$ ($\vec{b}_1$) and $\vec{b}_2^4$ ($\vec{b}_2$) denote the reciprocal primitive vectors for the 4×4 (1×1) supercell. $\Gamma^4$ ($\Gamma$), $K^4$ ($K$) and $K'^4$ ($K'$) are the high symmetry points for the 4×4 (1×1) supercell (c) Band structures of a 4×4 supercell of the pristine monolayer MoS$_2$. Red curve is used to denote the highest valence bands. Blue up-arrow and down-arrow denote respectively spin-up and spin-down electron.

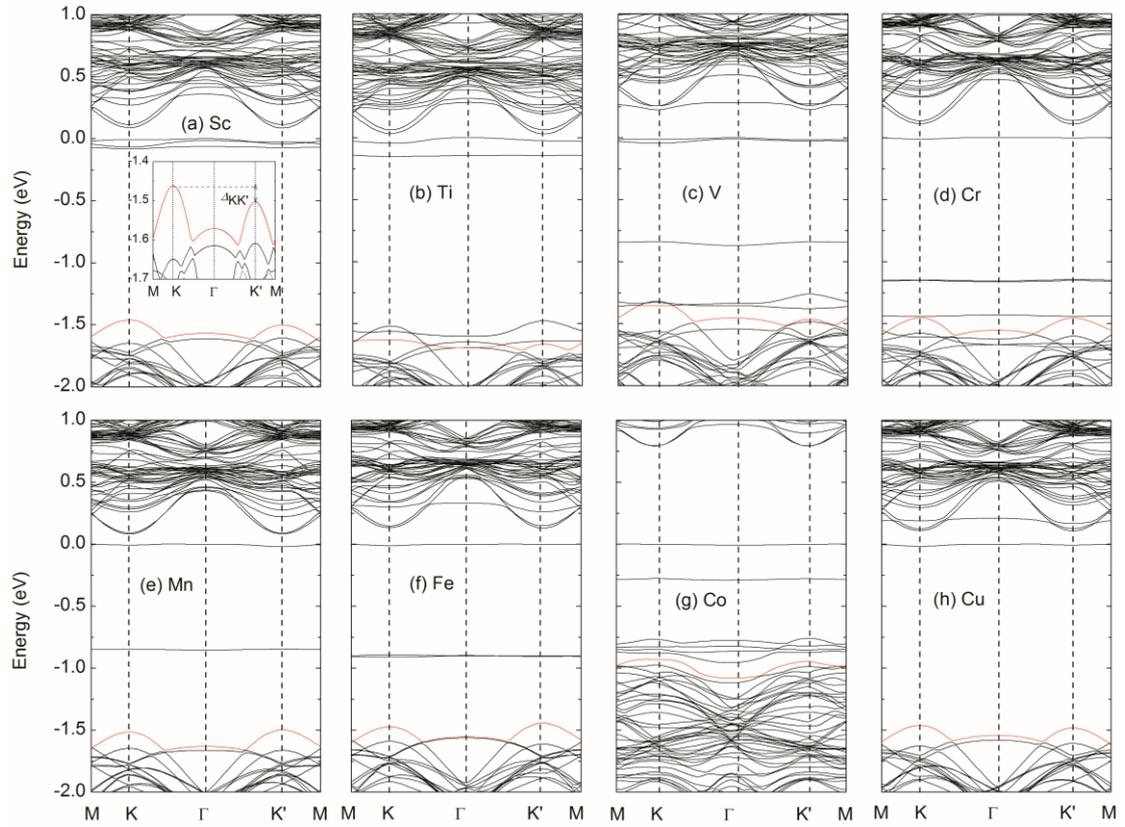

Figure.3. Band structures of a 4×4 supercell of MoS$_2$ monolayer adsorbed by Sc (a), Ti (b), V (c), Cr (d), Mn (e), Fe (f), Co (g) and Cu (h) atoms with the spin-orbit coupling. Fermi energy level is set to zero. Red curve are used to denote the highest valence bands. $\Delta_{KK'}$ denotes the energy difference between the highest valence band at K and K' points. The insertion in (a) is the magnification of the bands around the highest valence bands around K and K' points.

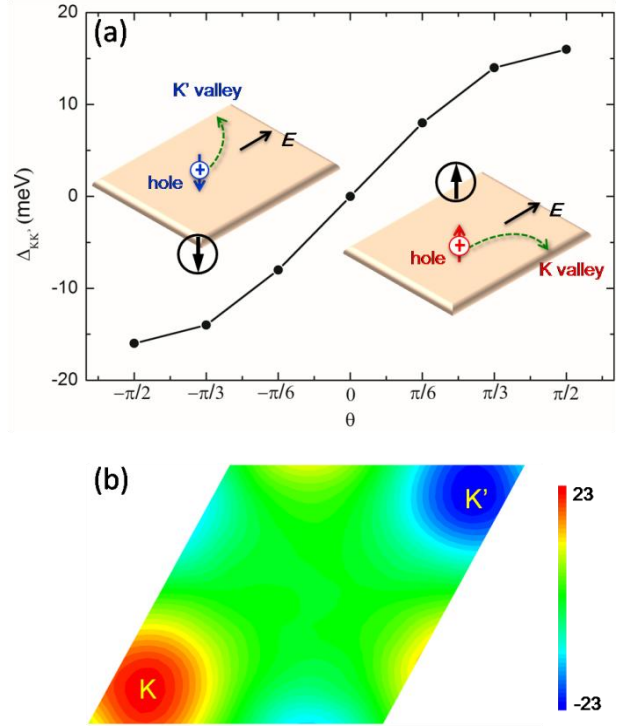

Figure.4. (a)The valley splitting $\Delta_{KK'}$ as a function of the magnetization direction. $\theta$ is the angle between the magnetic quantization axis and in-plane direction of monolayer. The insertion is the schematic depiction of anomalous Hall effects. The positive sign in circles denotes the hole. Holes with red up and blue down arrows denote up-spin one from K valley (right) and down-spin one from K' valley in (left), respectively. $E$ is the applied in-plane electric field. The black up and down arrows in circles show the magnetization direction with $\theta=\pi/2$ and $\theta=-\pi/2$, respectively. (b) $k$-resolved non-Abelian Berry curvature (units Å$^2$) of the valence bands occupied up to the MoS$_2$ gap of the system.

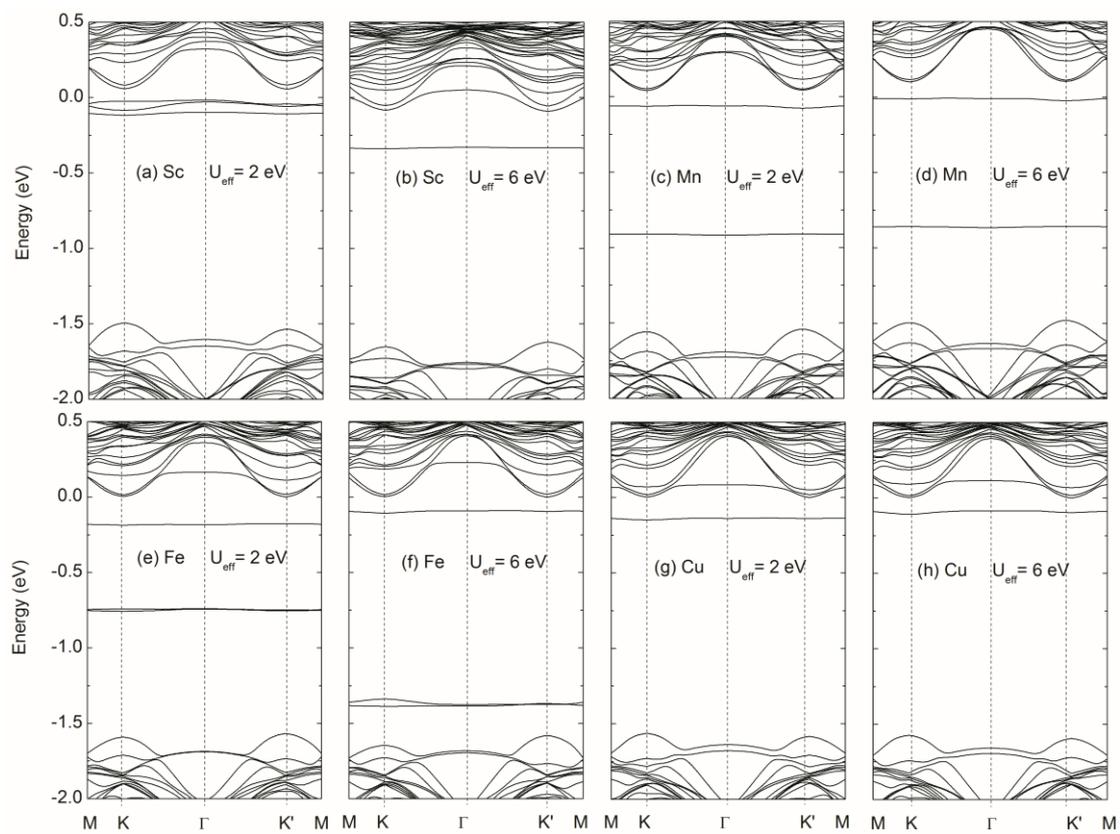

Figure.5. The band structures of a 4×4 supercell with a Sc, Mn, Fe or Cu atom adsorption when $U_{eff}$ = 2 and 6 eV, respectively.